# High-dimensional multiple imputation (HDMI) for partially observed confounders including natural language processing-derived auxiliary covariates

**Authors**: Janick Weberpals[1], Pamela A. Shaw[2], Kueiyu Joshua Lin[1], Richard Wyss[1], Joseph M Plasek[3], Li Zhou[3], Kerry Ngan[1], Thomas DeRamus[1], Sudha R. Raman[4], Bradley G. Hammill[4], Hana Lee[5], Sengwee Toh[6], John G. Connolly[6], Kimberly J. Dandreo[7], Fang Tian[8], Wei Liu[8], Jie Li[8], José J. Hernández-Muñoz[8], Sebastian Schneeweiss[1], Rishi J. Desai[1]

Author affiliations:

[1] Division of Pharmacoepidemiology and Pharmacoeconomics, Department of Medicine, Brigham and Women's Hospital, Harvard Medical School, Boston, MA, USA

[2] Biostatistics Division, Kaiser Permanente Washington Health Research Institute, Seattle, WA, USA

[3] Division of General Internal Medicine and Primary Care, Brigham and Women's Hospital, Harvard Medical School, Boston, MA, USA

[4] Department of Population Health Sciences, Duke University School of Medicine, Durham, NC, USA

[5] Office of Biostatistics, Center for Drug Evaluation and Research, US Food and Drug Administration, Silver Spring, MD, USA

[6] Department of Population Medicine, Harvard Medical School and Harvard Pilgrim Health Care Institute, Boston, MA, USA

7 Department of Population Medicine, Harvard Pilgrim Health Care Institute, Boston, MA, USA

8 Office of Surveillance and Epidemiology, Center for Drug Evaluation and Research, US Food and Drug Administration, Silver Spring, MD, USA

**Correspondence:**

Janick Weberpals, RPh, PhD

Division of Pharmacoepidemiology and Pharmacoeconomics,

Department of Medicine, Brigham and Women's Hospital, Harvard Medical School,

1620 Tremont Street, Suite 3030-R, Boston, MA 02120, USA

Phone: +1 617-278-0932

Fax: + 1 617-232-8602

Email: jweberpals@bwh.harvard.edu

**Article type:** Original Research Article

**Word count:** 3,806 words / 4000 words

**Tables:** 1

**Figures:** 6

**Supplementary material:** Supplementary tables and figures

**Short running title**: High-dimensional multiple imputation (HDMI)

**Funding Statement:** This project was supported by Task Order 75F40119F19002 under Master Agreement 75F40119D10037 from the US Food and Drug Administration (FDA). Additional funding was provided by NIH RO1LM013204.

**Competing Interests Statement:** The contents are those of the author(s) and do not necessarily represent the official views of, nor an endorsement, by FDA/HHS, or the U.S. Government. The FDA approved the study protocol, statistical analysis plan and reviewed and approved this manuscript. Coauthors from the FDA participated in the results interpretation and in the preparation and decision to submit the manuscript for publication. The FDA had no role in data collection, management, or analysis. Pamela Shaw is a named inventor on a patent licensed to Novartis by the University of Pennsylvania for an unrelated project. Joseph M Plasek reports personal fees from Credo Health outside the submitted work. Sengwee Toh serves as a consultant for Pfizer, Inc. and TriNetX, LLC for unrelated work. Dr. Schneeweiss is participating in investigator-initiated grants to the Brigham and Women's Hospital from Boehringer Ingelheim, Takeda, and UCB unrelated to the topic of this study. He owns equity in Aetion Inc., a software manufacturer. He is an advisor to Temedica GmbH, a patient-oriented data generation company. His interests were declared, reviewed, and approved by the Brigham and Women's Hospital in accordance with their institutional compliance policies. Dr. Desai reports serving as Principal Investigator on investigator-initiated grants to the Brigham and Women's Hospital from Novartis, Vertex, and Bristol-Myers-Squibb on unrelated projects. All remaining authors report no disclosures or conflicts of interest.

**Data sharing statement:** Patient-level data cannot be shared due to restriction under our data use agreement with CMS. However, simulated data with similar distributions as in the underlying eligible complete cohort used to generate the plasmode datasets can be generated using the `generate_data()` function in R which can be found under [https://gitlab-scm.partners.org/drugepi/bias_simulation_missing_data](https://gitlab-scm.partners.org/drugepi/bias_simulation_missing_data). The usage of this function and missingness generation is illustrated in [https://drugepi.gitlab-pages.partners.org/bias_simulation_missing_data/](https://drugepi.gitlab-pages.partners.org/bias_simulation_missing_data/).

**Analytic code sharing statement:** Patient cohorts used in this study were queried using SAS version 9.4 and all analyses were conducted in R version 4.3.2. Analytic code used in this study and detailed information on used R packages and versions can be found in in the `renv.lock` file at [https://gitlab-scm.partners.org/drugepi/hd-mi](https://gitlab-scm.partners.org/drugepi/hd-mi).

*Manuscript last updated: 2024-05-16 07:36:48.543017*

# Abstract

250 words/250 words

**Background**

Multiple imputation (MI) models can be improved by including auxiliary covariates (AC), but their performance in high-dimensional data is not well understood. We aimed to develop and compare high-dimensional MI (HDMI) approaches using structured and natural language processing (NLP)-derived AC in studies with partially observed confounders.

**Methods**

We conducted a plasmode simulation study using data from opioid vs. non-steroidal anti-inflammatory drug (NSAID) initiators (X) with observed serum creatinine labs (Z2) and time-to-acute kidney injury as outcome. We simulated 100 cohorts with a null treatment effect, including X, Z2, atrial fibrillation (U), and 13 other investigator-derived confounders (**Z1**) in the outcome generation. We then imposed missingness (MZ2) on 50% of Z2 measurements as a function of Z2 and U and created different HDMI candidate AC using structured and NLP-derived features. We mimicked scenarios where U was unobserved by omitting it from all AC candidate sets. Using LASSO, we data-adaptively selected HDMI covariates associated with Z2 and MZ2 for MI, and with U to include in propensity score models. The treatment effect was estimated following propensity score matching in MI datasets and we benchmarked HDMI approaches against a baseline imputation and complete case analysis with **Z1** only.

**Results**

HDMI using claims data showed the lowest bias (0.072). Combining claims and sentence embeddings led to an improvement in the efficiency displaying the lowest root-mean-squared-error (0.173) and coverage (94%). NLP-derived AC alone did not perform better than baseline MI.

**Conclusions**

HDMI approaches may decrease bias in studies with partially observed confounders where missingness depends on unobserved factors.

# Background

Administrative health insurance claims databases and electronic health records (EHRs) are important data sources for the conduct of real-world evidence (RWE) studies when they are suitable for the study question. While administrative health insurance claims databases have historically served as the foundation for many RWE studies, they are limited in their ability to record critical clinical confounders such as vital signs and laboratory results which are often captured only in EHR.[1] However, EHR typically exhibit high levels of missingness in such relevant confounders which challenges the statistical analysis of the data.

Multiple imputation (MI) is a common approach to address missing data and can lead to valid and efficient estimation of a desired target parameter if the missingness of a partially observed confounder does not depend on unobserved factors.[2] This *missing at random* (MAR) assumption may be inferred through a combination of a recently proposed principled approach to empirically assess the potential patterns and mechanisms for partially observed confounder (POC) data[3,4] and substantive expert knowledge. Nevertheless, the true underlying missingness mechanisms can never be inferred with absolute certainty from observed data.

To increase the efficient estimation of treatment effects in MI analyses and to improve the plausibility of the MAR assumption, it has been proposed to add auxiliary covariates (AC) to the imputation model.[5] AC are defined as covariates within the original data that are correlated with the POC and possibly related to the missingness of the POC, but are not part of the main analysis that estimates the treatment effect.[6] Several studies suggested that such inclusive imputation strategies lead to more efficient estimates and potentially reduce bias.[7,8] However, data-adaptive approaches to identify AC for MI models are not well understood and it was suggested that including too many AC can also lead to increased bias through inclusion of weakly correlated AC or via the inclusion of colliders.[9,10] In a recent simulation study, the use of least absolute shrinkage and selection operator (LASSO) models was identified as the best performing data-driven AC selection strategy overall.[8] However, this study focused on missing outcome rather than on missing confounder data and was limited to a lower-dimensional covariate space of possible candidate AC. Moreover, it is unknown to which extend features of unstructured clinical notes combined with state-of-the-art natural language processing (NLP) algorithms may augment the performance of imputation models.

In this plasmode simulation study, we aimed to develop and compare fully data-adaptive high-dimensional MI (HDMI) approaches. This approach includes a multi-step process that identifies, operationalizes and prioritizes candidate AC from available data dimensions such as structured administrative claims and unstructured EHR data (via NLP) for inclusion in MI models. We hypothesized that leveraging additional information of thousands of potential candidate AC may augment the performance of MI models by approximating otherwise unobserved information and hence making the MAR assumption more plausible. As part of this simulation, we thus investigated specifically whether HDMI can decrease bias and increase efficiency in RWE studies with partially observed confounders.

## Methods

A summary of the simulation setup is illustrated in Figure 1.

**Data sources and study population for plasmode data generation**

We first defined an empirical base cohort of patients with osteoarthritis who initiated opioids (exposure) or non-steroidal anti-inflammatory drugs (NSAIDs, comparator) as the foundation for the plasmode data generation. This cohort was identified using data from the Mass General Brigham Research Patient Data Registry EHR in Boston, linked with Medicare fee-for-service claims data spanning from 2007 to 2017. Medicare claims data included enrollment files, Part A (inpatient) claims, Part B (outpatient) claims and Part D (prescription drug coverage). The EHR data provided detailed clinical information including structured information such as vital signs and laboratory data as well as unstructured clinical notes and reports.

The index date was defined as the date of the first dispensing of either an opioid or NSAID in Medicare Part D file and patients were not eligible if they were exposed to either of the two drug classes in the 365 days prior to and including the date of the first dispensing. In addition, patients were required to have had at least 365 days of continuous enrollment in Medicare Part A+B+D and at least one EHR encounter in the 365 days prior to (including) the index date. All covariates were measured in the 365-day baseline window prior to (including) the index date. For structured claims codes, both inpatient and outpatient codes in any position were considered. We defined the outcome as the time from index date to acute kidney injury (AKI), where AKI was ascertained based on an algorithm published by Patel et al.[11]

Based on this empirical base cohort, we derived a sub-cohort with complete information on serum creatinine and further eligibility criteria (*eligible complete cohort,* Figure 2.). Serum creatinine (referred to as $Z2$ for the remainder of the manuscript[2]) is a strong prognostic factor for AKI[12] and was used in this simulation as the confounder of interest on which we assessed the impact of missing data on.

**Plasmode data generation**

The plasmode data generation was described earlier in detail by Franklin et al.[13] We chose a plasmode simulation because it enabled us to realistically assess the effect of identifying and including auxiliary covariates in MI analyses by retaining the complexity and correlation of covariates observed in our real-world empirical cohort while allowing the investigator to determine the confounding and true effect size for the treatment exposure.

To that end, we first defined a covariate vector of 13 investigator-defined prognostic factors (henceforth referred to as **Z1**) and a patient's history of atrial fibrillation ($U$, investigator-derived variable defined as the presence of any ICD-9-CM 427.31 or ICD-10-CM I48.0, I48.1,

I48.2, I48.91 code observed in any position of an inpatient or outpatient claim in the baseline window) which were fitted along with the exposure ($X$) and $Z2$ using a multivariable Cox proportional hazards regression to model the empirical outcome and censoring function in the eligible complete cohort (**Supplementary Table 1**). Using Breslow estimates[14], we generated the model-estimated event and censoring times for each patient under the investigator-defined null treatment effect for $X$ (Hazard ratio $[\text{HR}]_{\text{true}} = 1$) and the patient's own covariate values. In the final stage, the baseline event-free survival function was adjusted to retain the same event rate as was observed in the eligible complete cohort, and a final time-to-event outcome was generated for each patient as the minimum of the predicted event time and censoring time (**Supplementary Figures 1 and 2**). The corresponding directed acyclic graph for the eligible complete cohort (c-DAG) is illustrated in Figure 3 a).

**Missingness generation**

To impose missingness on $Z2$, we used a multivariate *amputation* procedure to simulate a missing-not-at-random (MNAR) mechanism where the missingness probability was dependent on unobserved factors (m-DAG, Figure 3 b) and a missingness proportion of 50%.[15] To that end, we first defined a weighted sum score ($wss$) which was a weighted linear combination of a patient's ($i$) values of $Z2$ and $U$ (Equation 1).

$$wss_i = 0.8 \cdot U_i + 0.2 \cdot Z2_i \tag{1}$$

To mimic a scenario where $U$ was unmeasured for imputation and propensity score models, we subsequently omitted $U$ from the set of available candidate variables (*note that single ICD-9-CM or ICD-10-CM codes that were used to derive $U$ or related natural language in unstructured notes were not explicitly omitted from the empirical high-dimensional candidate covariates*).

The weighted sum score was additionally scaled by subtracting the mean of $wss_i$ from $wss_i$ and dividing by the standard deviation of $wss$.

$$wss_{i,scaled} = \frac{wss_i - \overline{wss}}{sd_{wss}} \tag{2}$$

Next, we mapped the $wss_{i,scaled}$ to discrete missingness probabilities by dividing the patients into four equally-sized quantiles based on their $wss_{i,scaled}$ and assigning odds values of each quantile group for having an unobserved $Z2$ value.[16] In this simulation, patients in the first quantile were assigned an odds = 1, patients in the second quantile and odds = 2, patients in the third quantile an odds = 3 and patients in the fourth quantile an odds = 4. That is, patients in the highest quantile had a four-fold increased odds of having an unobserved $Z2$ compared to patients in the first quantile (with corresponding missingness indicator notation $MZ2 = 1$ if $Z2$ is missing and $MZ2 = 0$ if $Z2$ is observed). This translates into a clinical scenario where sicker patients with higher serum creatinine levels and a history of atrial fibrillation are more likely to have a missing serum creatinine measurement.[17]

### HDMI algorithm

We hypothesized that by identifying and prioritizing AC from high-dimensional covariate vectors using a multi-step HDMI approach (Figure 4), including such empirical correlates of $Z2$ and $MZ2$ in the imputation step may lead to more efficient and less biased estimates.

**Step 1 - identification of covariate dimensions:** The first step involved the identification of the available data sources. Administrative insurance claims typically contain multiple different code dimensions such as diagnostic codes (ICD-9/10-CM), procedural codes (Current Procedural Terminology [CPT-4], Healthcare Common procedure Coding System [HCPCS]) and information on prescribed drugs. In this study, unstructured clinical notes were additionally available through linkage with EHR data of eligible patients.

**Step 2 - covariate operationalization:** To operationalize the identified covariate dimensions, codes from structured claims were binarized to indicate the presence of a specific code in the 365 days before (and including) the index date. This resulted in a total of 28,874 empirical candidate covariates from claims codes (*HDMI claims*). For EHR-based clinical notes, two NLP algorithms were employed to operationalize features from unstructured free-text. First, we applied a bag-of-words *Ngram* model that generated binary candidate covariates indicating the presence of a unique word (unigram) across all recorded physician notes for a patient in the baseline window.[18] The frequency or order of words within a sentence were not considered for this approach and we excluded stop words that conveyed only little semantic meaning like "a", "in", etc. Overall, this resulted in 19,993 empirical binary unigram covariates (*HDMI unigram*). We additionally considered sentence embeddings derived from *Bidirectional Encoder Representations from Transformers* (BERT) models.[19,20] In brief, this NLP algorithm first tokenizes sentences from a physician note into a numerical vector of word pieces which is used as input into a multi-layer bidirectional transformer model.[21] The transformer processes the tokens in both directions (bidirectional) to capture the context of each token and hence encodes the semantic meanings of a sentence in a 128-dimensional vector. The sentence embeddings in this study were generated using John Snow Lab's open-source Spark NLP pre-trained models available at https://github.com/JohnSnowLabs/nlu. Since patients could have multiple sentence embeddings, mean pooling was applied to arrive at a 128-dimensional embedding per patient (*HDMI sentence*). Lastly, we also considered two vectors that blended covariate candidate vectors from claims with unigram (HDMI claims + unigram) and sentence embeddings (HDMI claims + sentence), respectively.

**Step 3 - prevalence filter**: For empirical binary covariate candidates (HDMI claims and HDMI unigrams), we added a filtering step to exclude covariates with a prevalence $< 1\%$.

Covariates with such a low prevalence are unlikely to provide a substantial amount of information and filtering out such codes significantly improves the computational performance of subsequent steps.

**Step 4 - empirical covariate prioritization**: To data-adaptively prioritize **Z1** and HDMI candidate covariates for the imputation (step 4) and propensity score models (step 5), we fitted three independent regularized regressions. Covariates for the imputation model were selected using two penalized LASSO regression models that independently minimized the prediction error for $Z2$ in the complete cases ($LASSO_{Z2}$) and $MZ2$ in all patients ($LASSO_{MZ2}$).[22,23]

For both models, $X$ and $Y$ were forced into the model.[24] To account for the introduced unmeasured confounding, we fitted a third outcome-adaptive LASSO model ($LASSO_{PS}$) that prioritized covariates that minimized the deviance (-2 log partial likelihood) for the outcome using a regularized Cox proportional hazard model for right-censored time-to-event outcomes (with exposure $X$ being forced into the model).[25,26]

For all models, the penalization factor $\lambda$ was determined using 5-fold cross validation.

**Step 5 - imputation**: For each HDMI approach, $m = 10$ imputed datasets were created using a predictive mean matching algorithm for $Z2$.[27] The final predictor matrix for the imputation model comprised the intersecting covariates identified in both $LASSO_{Z2}$ and $LASSO_{MZ2}$ as $Imputation_{\text{predictor matrix}} = LASSO_{Z2} \cap LASSO_{MZ2}$.

**Step 6 - propensity score and outcome model**: To estimate the treatment effects for $X$ using propensity score matching, we applied the "within" approach.[28,29] That is, propensity score matching and the estimation of the treatment effect are performed in each imputed dataset separately and resulting treatment effect estimates are combined using Rubin's rule. In this study this was implemented by fitting a propensity score model in each imputed dataset using the covariates identified in $LASSO_{PS}$. Patients were then matched in a 1:1 ratio using a nearest neighbor matching algorithm with a 0.2 caliper of the propensity score without replacement. For each matched dataset, a Cox proportional hazards regression model was fit to estimate the marginal average treatment effect in the matched population. Confidence intervals were estimated using cluster-robust standard errors.[30] The final treatment effect estimates for $X$ were then combined using Rubin's rule.[31]

**HDMI model comparisons and simulation performance metrics**

The overall aim of the simulation was to compare the performance of different structured, unstructured and combinations of HDMI candidate AC. To benchmark the HDMI models, we additionally considered an unadjusted analysis, a complete case analysis and a baseline model, latter of which could only select from the investigator-derived **Z1** covariates in step 4, but not from any of the empirical HDMI data dimensions. A summary on all compared models and corresponding candidate covariates is provided in **Table 1** and **Supplementary Figure 3**.

The generation of missing data and application of each HDMI model was repeated in each of the 100 simulated plasmode cohorts. To evaluate the different models, we computed the root mean squared error (RMSE) and bias as our primary metrics of interest, as well as variance and coverage of the nominal 95% confidence interval (CI) to evaluate how well the different models were able to recover the true HR for the treatment effect (opioids vs. NSAIDs, **Supplementary Table 2**).

### Statistical software and reproducibility

All patient cohorts used in this study were created using SAS version 9.4 and all analyses were conducted in R version 4.3.2. Detailed information on used R packages, versions and code used in this study is available at https://gitlab-scm.partners.org/drugepi/hd-mi.

# Results

Applying all eligibility criteria led to a final population of 5,949 patients with an overall crude incidence rate for AKI of 36.35/1,000 person-years (**Supplementary Figure 4**).

Figure 5 illustrates the distribution of number of covariates selected by the LASSO models for a) the imputation model and b) the propensity score model across all 100 simulated plasmode datasets. Given the restriction of using only intersecting covariates that were independent predictors of both $Z2$ and $MZ2$ led to a median (interquartile range) between 4 (2-6, unigrams) and 13 (8-16, HDMI sentence embeddings) covariates that were included in the respective imputation models in addition to $X$, $Y$ and $Z2$. The outcome-based LASSO selected a median between 9 (7-10, complete case analysis) and 28 (20-40, HDMI claims + sentence) covariates for the propensity score model.

The main results are summarized in Figure 6. Overall, the HDMI claims model showed the lowest bias (0.072 [95% CI 0.04-0.104]), with a RMSE of 0.179 (0.154-0.201) and coverage of 93% (88%-98%). Comparing the two different NLP-based models, the sentence embedding model performed better than the unigram approach (RMSE 0.206 [0.176-0.232] versus 0.233 [0.195-0.265]) but led to only marginal improvement when compared to the baseline model (RMSE 0.210 [0.179-0.238]). However, combining sentence embeddings with claims led to slightly more efficient estimates than with claims alone through a decrease of variance (0.024) and improved coverage (94% [0.89%-0.99%]) and hence showed the lowest overall RMSE (0.173 [0.151-0.192]). The complete case analysis exhibited comparable bias to the baseline model (0.127 [0.074-0.18] versus 0.132 [0.099-0.164]) but significantly higher variance (0.072 versus 0.027) which translated to the highest RMSE (0.297 [0.245-0.341]). Expectedly, the unadjusted model showed the largest bias across all models (0.214 [0.181-0.248]).

# Discussion

In this study, we developed and compared different fully data-adaptive HDMI models, including NLP-derived auxiliary covariates, to address partially observed confounders in the presence of a MNAR scenario and unmeasured confounding. Across all compared models, the HDMI claims model led to the lowest observed bias while the combination of claims and sentence embeddings showed the most efficient treatment effect estimation and bias-variance trade-off.

The benefit of including AC in MI has been explained by the increase in statistical efficiency in the treatment effect estimation.[5] However, not much is known about the use of high-dimensional AC to also decrease bias in MI analyses, which is highly relevant in scenarios where missingness cannot be fully explained by observed covariates. Betancur et al.[2] introduced canonical causal diagrams for the handling of multivariate missing data and assessed under which scenarios a target parameter is recoverable, that is, a statistical quantity can be consistently estimated as a function of the available data distribution. They concluded that when no variable in the underlying data generating mechanism causes its own missingness, joint distributions and target parameters are recoverable.

Our simulation study explored if the inclusion of high-dimensional AC can simultaneously help to not only increase efficiency but also decrease bias in the presence of a multivariate MNAR mechanism by approximating the joint distribution of unobserved factors through correlated AC. The HDMI claims model was observed to lead to a decrease in bias when compared to standard approaches (complete case analysis or the baseline MI model that included only investigator-derived covariates). This may be related to several factors. High-dimensional propensity score models follow the same idea of including proxy covariates to account for unobserved factors and have consistently shown to decrease bias in the presence of unmeasured confounding.[32] Given the comprehensive translation of medical concepts into structured medical diagnosis, procedure and treatment codes, it is likely that stronger correlates of $U$ or $Z2$ may be selected by the LASSO models, especially since we only selected predictors of both $Z2$ and $MZ2$.

We found that considering NLP-based covariates in addition to claims-based AC did not improve HDMI performance. Claims data have the advantage that they record the entirety of a patient's medical history and hence ensure data continuity.[33] It has been shown that gaps in the longitudinal data capture can have detrimental consequences for the measurement of important variables when studying treatments effects and may lead to differential misclassification and information bias between patients with high versus low data continuity.[34] In this study, NLP-derived features came from tertiary single-institutional EHR data which may provide more in-depth clinical information but likely missed important medical information when patients received care at a different institution outside the Mass General Brigham network. Post-hoc analyses confirmed that a large number of patients in the eligible complete cohort showed only low EHR data continuity with less than a mean of 50% of encounters captured in the EHR data (**Supplementary Figure 5**).[33] Hence, this low coverage of EHR likely jeopardized the ability of NLP models to find all meaningful AC. Comparing the two different NLP approaches in spite of data discontinuity, the sentence embeddings model showed better performance. This may be due to the more contextual modeling of natural language in BERT models which enable the learning of semantic meanings of sentences rather than the simpler one-hot-encoding of words in bag-of-word approaches.

Interestingly, the number of final selected covariates for imputation models was rather low across all models. This is related to the restrictive condition of including only strong predictors of both $Z2$ and $MZ2$. We observed that each individual LASSO model selected a significantly higher amount of covariates with only modest overlap in intersecting covariates. This overlap may reflect only the most important covariates needed to sufficiently impute the partially observed $Z2$ covariate and may differ from study to study based on the amount of partially observed confounders, the correlation between covariates and the proportion missingness. While in the literature there are arguments for both inclusive and restrictive covariate selection strategies[7,8,10,35], with the availability of high-dimensional data, the more restrictive approach may be desirable since including too many covariates may often lead to convergence problems or unreasonably long computation times without any additional information gain, e.g., due to collinearity, perfect prediction, or weakly correlated covariates.

## Strengths and limitations

One potential limitation of this study is that we based our assumptions on a single MNAR missingness mechanism where missingness was imposed by unobserved predictor covariates. While there is an almost indefinite number of potential missingness scenarios, we decided on this missingness generating mechanisms since it represents a plausible clinical scenario as it could be expected from EHR. In other realistic scenarios there may be more unobserved factors in which case it is uncertain if the results of this simulation study would generalize. Moreover, our focus was on assessing the benefit of adding HDMI AC in more extreme and complex MNAR missingness scenarios since we recognize that AC may not contribute much in terms of bias reduction to the imputation of missing data in the case of MAR mechanisms. We further focused on partially observed confounder data, rather than AC selection strategies for missing exposures or outcomes for which data has been published previously.[8,10]

There may have also been alternative approaches for covariate prioritization than LASSO (step 4) such as using multiple univariate pairwise correlations (as in the `quickpred` function implemented in the mice package[36]) or more complex dimensionality reduction algorithms like autoencoder neural networks.[37] We decided to choose LASSO as it has been shown to consistently perform well and is highly interpretable in multivariate covariate prioritization tasks.[8,18,38] In addition, alternative methods, which are computationally not optimized, were prohibitively slow for the large amount of covariates considered in this simulation study.

Lastly, the eligibility criteria for the empirical complete cohort used for simulating plasmode datasets spanned both ICD-9-CM and ICD-10-CM eras which may affect the performance of claims-based HDMI models. However, a recent study[39] has shown that there is only meaningful bias in situations where the exposure distribution between the two compared patient cohorts is disproportionate between ICD-9 and ICD-10 eras which can then act as instrument-like covariates. This was not observed in our empirical cohort (**Supplementary Table 3**).

A strength of our study is the inclusion and detailed comparison of structured and unstructured high-dimensional covariate spaces to derive AC. We also considered a rather complex, yet realistic multivariate MNAR missingness generating mechanism which represents a clinically plausible scenario. Moreover, we employed a plasmode simulation approach since this method retains more complex covariate correlations inherent to real-world data and hence may lead to more realistic simulations compared to de novo simulations.[13,40].

## Conclusions

In this simulation study, HDMI based on claims data was the approach that decreased bias the most and combining HDMI claims with NLP-derived sentence embeddings led to the best overall bias-variance trade off and most efficient treatment effect estimation. These HDMI models may be considered in studies with partially observed confounders and missingness scenarios where missingness depends on unobservable factors.

## Tables

Table 1. Comparison of covariate candidates by model.

| Model | Candidate covariates[a] | # candidate covariates | Encoding |
|---|---|---|---|
| Unadjusted | - | - | - |
| Complete case | Investigator-derived (Z1) | 13 | Mixed |
| Baseline model | Investigator-derived (Z1) | 13 | Mixed |
| HDMI claims | Medicare claims | 28,874 (claims) | Binary |
| HDMI unigram | NLP unigram | 19,993 (unigram) | Binary |
| HDMI sentence | NLP BERT sentence embeddings | 128 (sentence embeddings) | Continuous |
| HDMI claims + unigram | Medicare claims + NLP unigram | 28,874 (claims) + 19,993 (unigram) | Binary |
| HDMI claims + sentence | Medicare claims + NLP BERT sentence embeddings | 28,874 (claims) + 128 (sentence embeddings) | Mixed |

Abbreviations: BERT = Bidirectional encoder representations from transformers, HDMI = High-dimensional multiple imputation, Z1 = investigator-derived covariates used in outcome-generation model: Age at index date, No. of ED visits, No. of distinct prescriptions, Atrial fibrillation, Flu vaccine, Foot ulcer, Glaucoma or cataract, Ischemic stroke, H2 Receptor Antagonist, ACE-Inhibitors, ARBs, Statins, Spironolocatone

[a] All HDMI models are also allowed to select from the 13 investigator-derived (Z1) covariates as candidate covariates.

# Figures

Figure 1: Illustration of plasmode simulation setup.

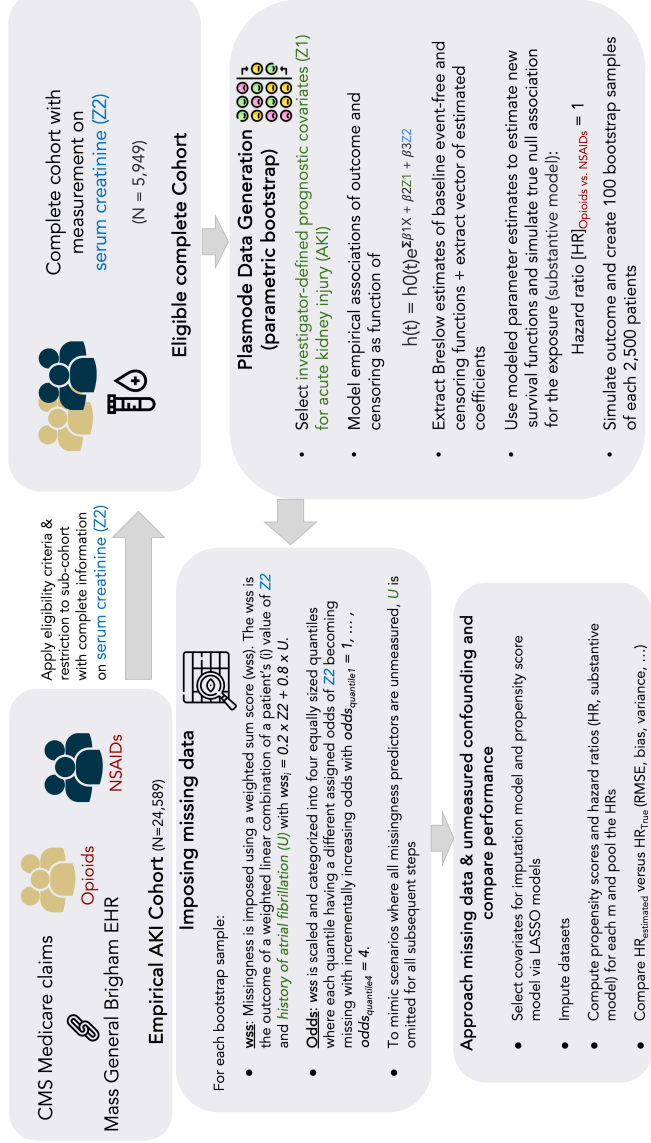

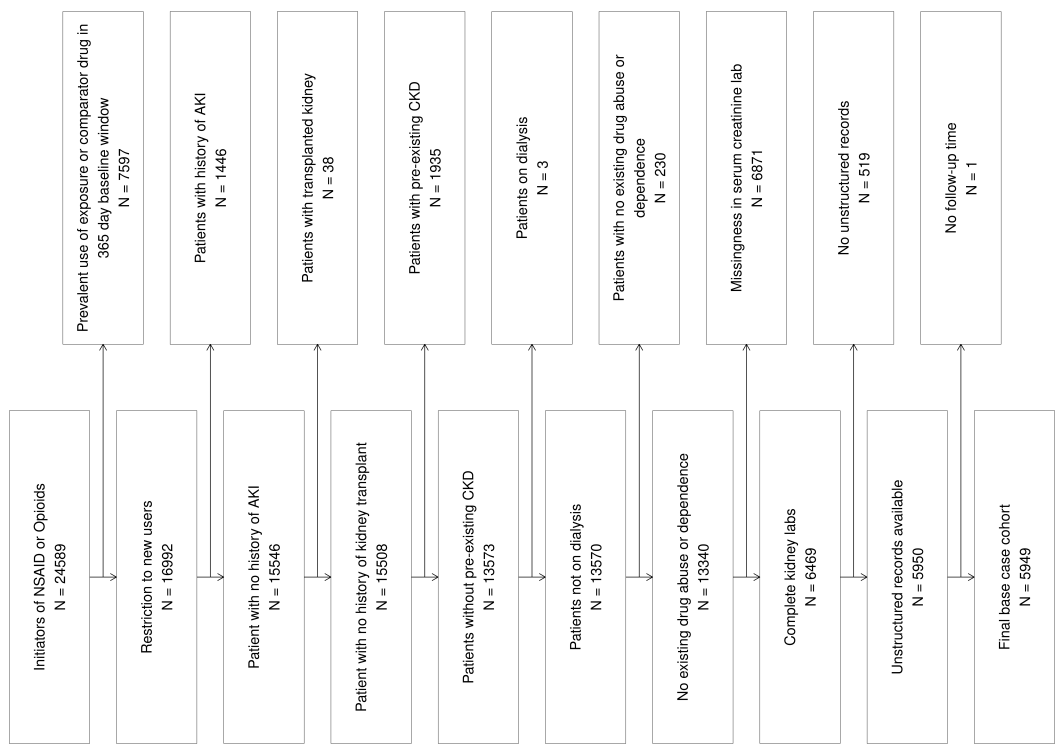

Figure 2: Attrition of eligible empirical complete cohort used as a basis to create plasmode datasets.

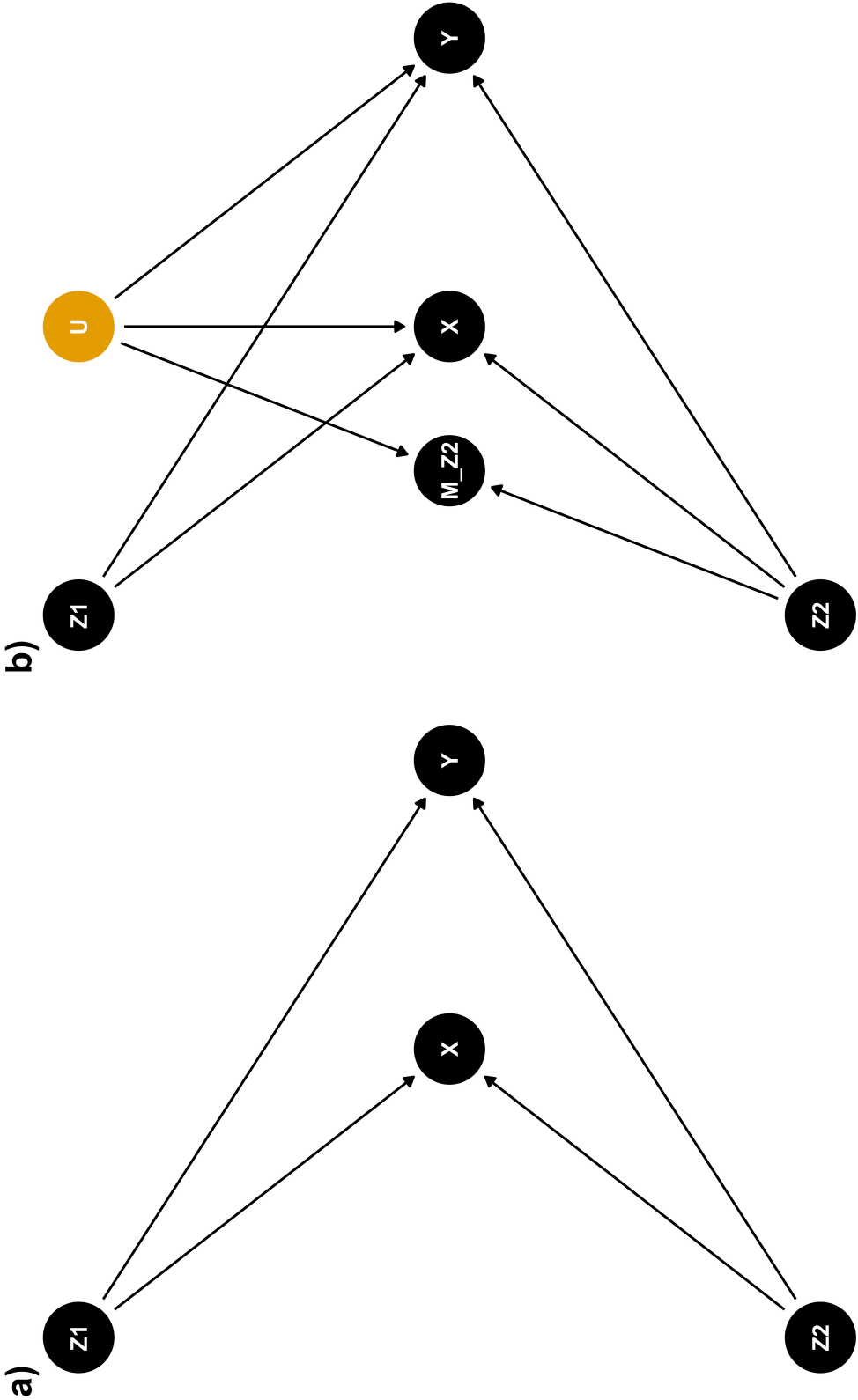

Figure 3: Directed acyclic graphs illustrating the (a) complete data (c-DAG) and (b) missingness (m-DAG) generating mechanisms. M_Z2 = missingness indicator for Z2, U = History of atrial fibrillation (unobserved for imputation and propensity score models), X = Exposure, Z1 = fully observed investigator-derived covariates, Z2 = serum creatinine, Y = Time to acute kidney injury

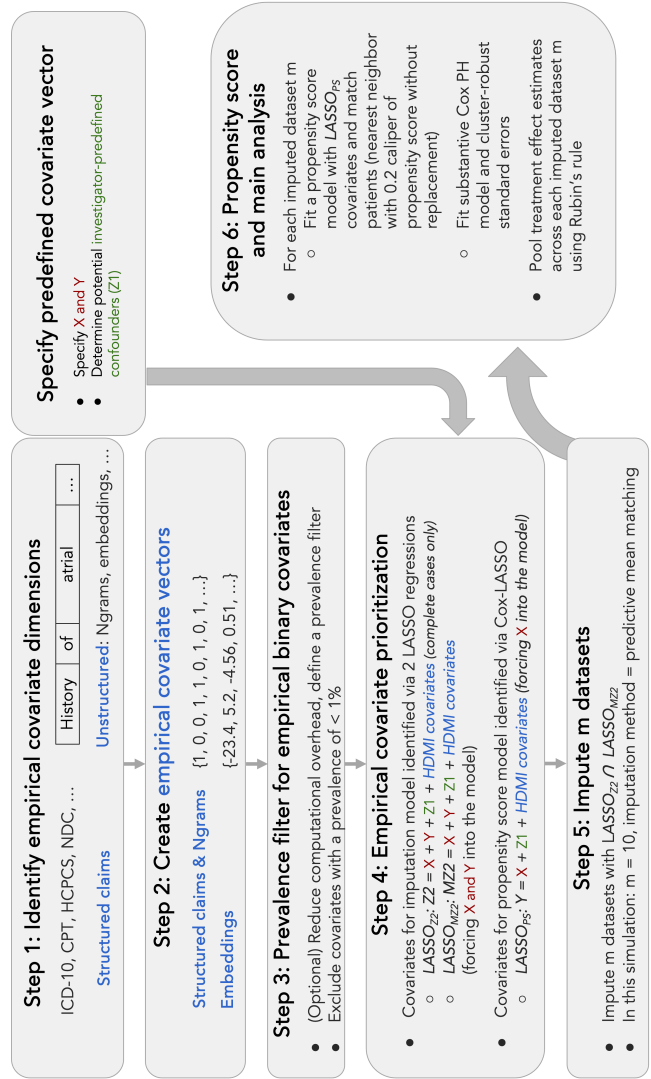

Figure 4: Illustration of high-dimensional multiple imputation (HDMI) algorithm.

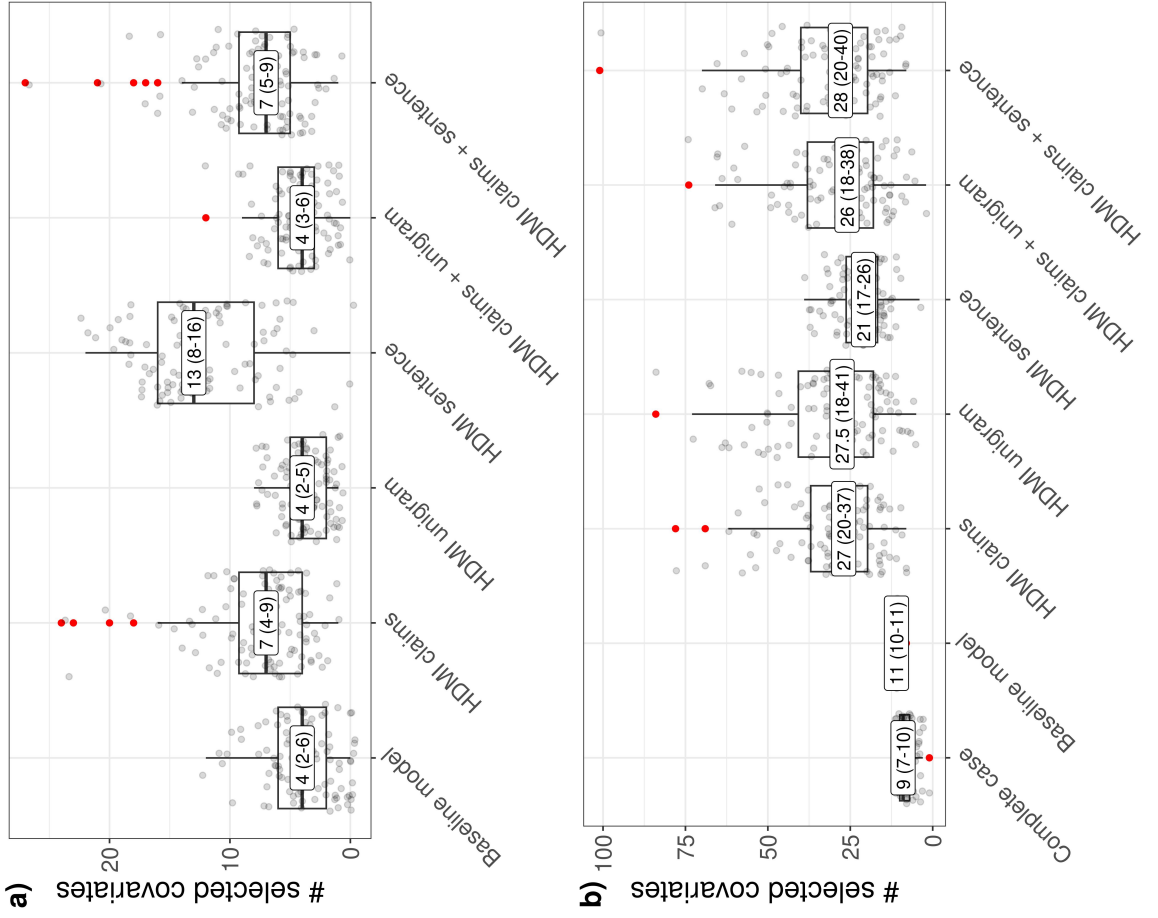

Figure 5: Distribution (median, interquartile range) of the number of covariates selected for a) the baseline and HDMI imputation models (X, Y and Z2 are not considered since those are always included) and b) the propensity score model (Z2 is not considered since it is always included).

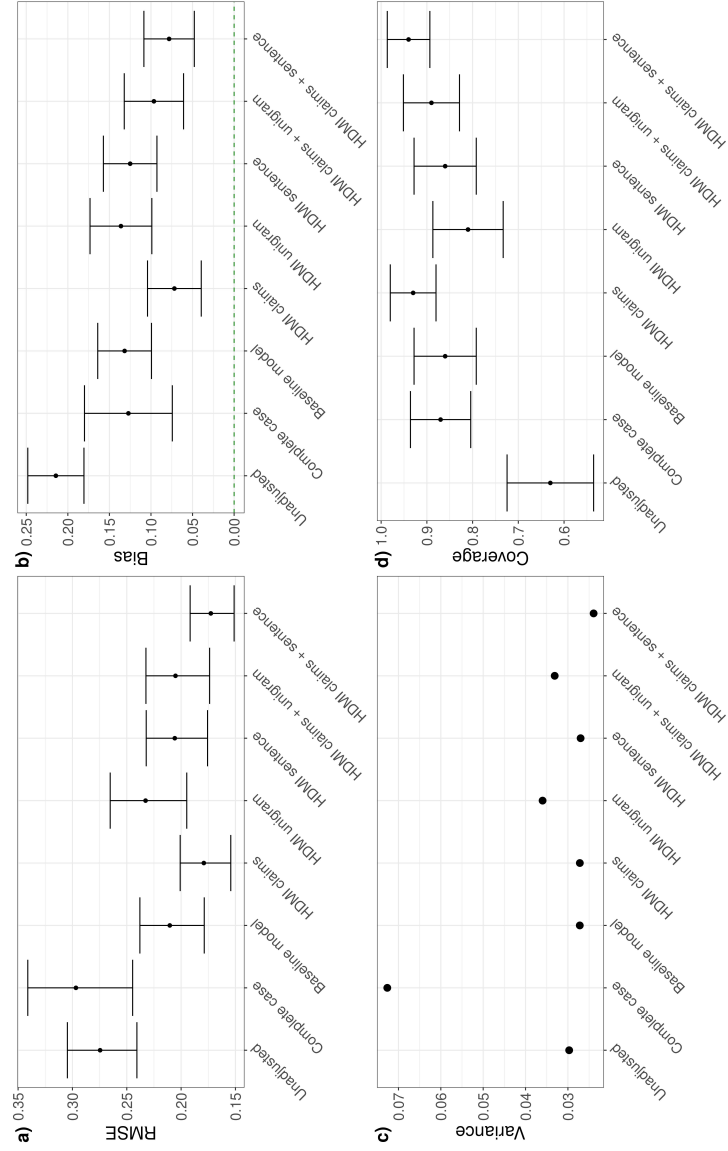

Figure 6: Main results illustrating the a) root-mean-squared-error (RMSE), b) bias, c) variance and d) coverage of the nominal 95% confidence interval (CI) between analytical methods to account for partially observed serum creatinine (Z2) measurements and unmeasured confounding.

# Supplementary Material

**High-dimensional multiple imputation (HDMI) for partially observed confounders including natural language processing-derived auxiliary covariates**

**Authors**: Janick Weberpals[1], Pamela A. Shaw[2], Kueiyu Joshua Lin[1], Richard Wyss[1], Joseph M Plasek[3], Li Zhou[3], Kerry Ngan[1], Thomas DeRamus[1], Sudha R. Raman[4], Bradley G. Hammill[4], Hana Lee[5], Darren Toh[6], John G. Connolly[6], Kimberly J. Dandreo[7], Fang Tian[8], Wei Liu[8], Jie Li[8], José J. Hernández-Muñoz[8], Sebastian Schneeweiss[1], Rishi J. Desai[1]

Author affiliations:

[1] Division of Pharmacoepidemiology and Pharmacoeconomics, Department of Medicine, Brigham and Women's Hospital, Harvard Medical School, Boston, MA, USA

[2] Biostatistics Division, Kaiser Permanente Washington Health Research Institute, Seattle, WA, USA

[3] Division of General Internal Medicine and Primary Care, Brigham and Women's Hospital, Harvard Medical School, Boston, MA, USA

[4] Department of Population Health Sciences, Duke University School of Medicine, Durham, NC, USA

[5] Office of Biostatistics, Center for Drug Evaluation and Research, US Food and Drug Administration, Silver Spring, MD, USA

[6] Department of Population Medicine, Harvard Medical School and Harvard Pilgrim Health Care

Institute, Boston, MA, USA

7 Department of Population Medicine, Harvard Pilgrim Health Care Institute, Boston, MA, USA

8 Office of Surveillance and Epidemiology, Center for Drug Evaluation and Research, US Food and Drug Administration, Silver Spring, MD, USA

*Supplementary Material last updated: 2024-04-26*

# 1 Supplementary Tables

Supplementary Table 1: Overview on all covariates involved in the substantive model along with exposure and serum creatinine.

| Variable | Label | Type |
| --- | --- | --- |
| dem_age | Age at index date | continuous |
| c_ed | No. of ED visits | continuous |
| c_gnrc_cnt | No. of distinct prescriptions | continuous |
| c_atrial_fibrillation | Atrial fibrillation | binary |
| c_flu_vaccine | Flu vaccine | binary |
| c_foot_ulcer | Foot ulcer | binary |
| c_glaucoma_or_cataract | Glaucoma or cataract | binary |
| c_ischemic_stroke | Ischemic stroke | binary |
| c_h2ra | H2 Receptor Antagonist | binary |
| c_acei | ACE-Inhibitors | binary |
| c_arb | ARBs | binary |
| c_statin | Statins | binary |
| c_spironolocatone | Spironolocatone | binary |

## 1.1 Imputation performance assessment

To evaluate the performance of the different imputation approaches, we derived different metrics comparing the difference in the resulting treatment effect estimate/hazard ratio (HR) and the true treatment effect estimate of $HR_{true} = 1$. These metrics included the root mean squared error (RMSE), bias, variance and coverage of the nominal 95% confidence interval as described by Morris et al.[1] The summary statistics and performance metrics were derived using the `rsimsum` R package (version 0.11.3).[2]

Supplementary Table 2: Details on performance measures used to evaluate and compare between imputation approaches.

| Metric | Formula |
|---|---|
| RMSE | $\sqrt{\frac{1}{n_{sim}} \sum_{i=1}^{n_{sim}} (HR_i - 0)^2}$ |
| Bias | $\frac{1}{n_{sim}} \sum_{i=1}^{n_{sim}} (HR_i - 0)$ |
| Variance | $\frac{1}{n_{sim}-1} \sum_{i=1}^{n_{sim}} (HR_i - \overline{HR})^2$ |
| Coverage | $\frac{1}{n_{sim}} \sum_{i=1}^{n_{sim}} \left( HR_{\text{true}} \geq HR_{\text{lower},i} \& HR_{\text{true}} \leq HR_{\text{upper},i} \right)$ |

Coverage = Coverage of the estimated nominal 95% confidence interval, $HR_i$ = estimated hazard ratio in iteration $i$, $HR_{\text{lower/upper},i}$ = Lower/upper 95% confidence interval limit for estimated hazard ratio in iteration $i$, $HR_{\text{true}}$ = True hazard ratio, RMSE = Root mean squared error

Supplementary Table 3: Patient characteristics of eligible complete cohort stratified by exposure.

| Patient characteristic | Total N = 5949[1] | Treatment received | | Difference[2] |
| --- | --- | --- | --- | --- |
| | | NSAIDs N = 1876 (31.5%)[1] | Opioids N = 4073 (68.5%)[1] | |
| Age at index date | 74 (70, 81) | 74 (69, 79) | 75 (70, 82) | -0.20 |
| No. of ED visits | 1 (0, 2) | 0 (0, 1) | 1 (0, 2) | -0.12 |
| No. of distinct prescriptions | 10 (7, 13) | 9 (6, 13) | 10 (7, 14) | -0.14 |
| Atrial fibrillation | 1,052 (18%) | 177 (9.4%) | 875 (21%) | -0.34 |
| Flu vaccine | 3,804 (64%) | 1,198 (64%) | 2,606 (64%) | 0.00 |
| Foot ulcer | 232 (3.9%) | 45 (2.4%) | 187 (4.6%) | -0.12 |
| Glaucoma or cataract | 3,142 (53%) | 1,015 (54%) | 2,127 (52%) | 0.04 |
| Ischemic stroke | 715 (12%) | 175 (9.3%) | 540 (13%) | -0.12 |
| Concomitant medication | | | | |
| H2 Receptor Antagonist | 426 (7.2%) | 143 (7.6%) | 283 (6.9%) | 0.03 |
| ACE-Inhibitors | 1,891 (32%) | 587 (31%) | 1,304 (32%) | -0.02 |
| ARBs | 1,008 (17%) | 290 (15%) | 718 (18%) | -0.06 |
| Statins | 3,438 (58%) | 1,127 (60%) | 2,311 (57%) | 0.07 |
| Spironolocatone | 125 (2.1%) | 30 (1.6%) | 95 (2.3%) | -0.05 |
| Calendar year of index date | | | | 0.07 |
| <2016 | 4,274 (72%) | 1,340 (71%) | 2,934 (72%) | |
| 2016 | 734 (12%) | 212 (11%) | 522 (13%) | |
| 2017 | 941 (16%) | 324 (17%) | 617 (15%) | |
| Serum creatinine level [mg/dL] | 0.83 (0.70, 1.00) | 0.85 (0.71, 1.00) | 0.82 (0.70, 0.99) | 0.10 |

[1]Median (IQR); n (%)
[2]Standardized Mean Difference

# 2 Supplementary Figures

Supplementary Figure 1: Comparison of log odds for initiating exposure in original and (averaged) plasmode dataset.

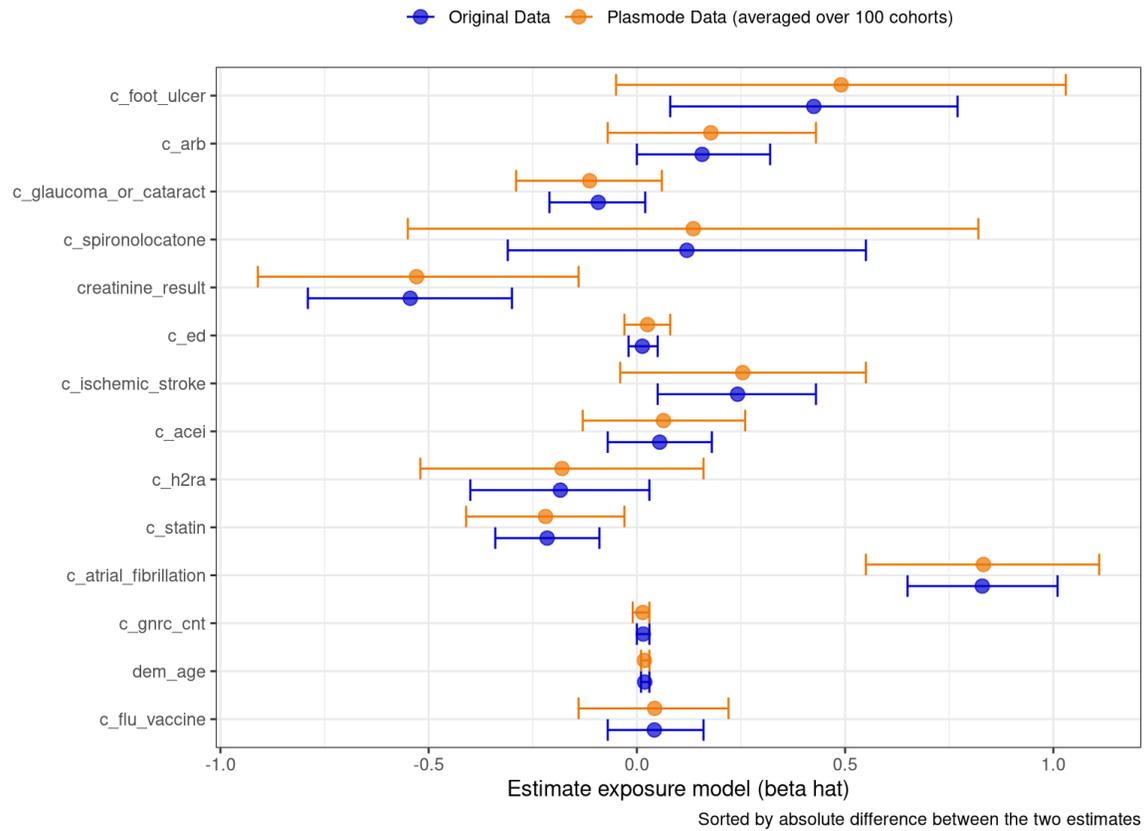

Supplementary Figure 2: Comparison of log hazard ratios for the outcome in original and (averaged) plasmode dataset.

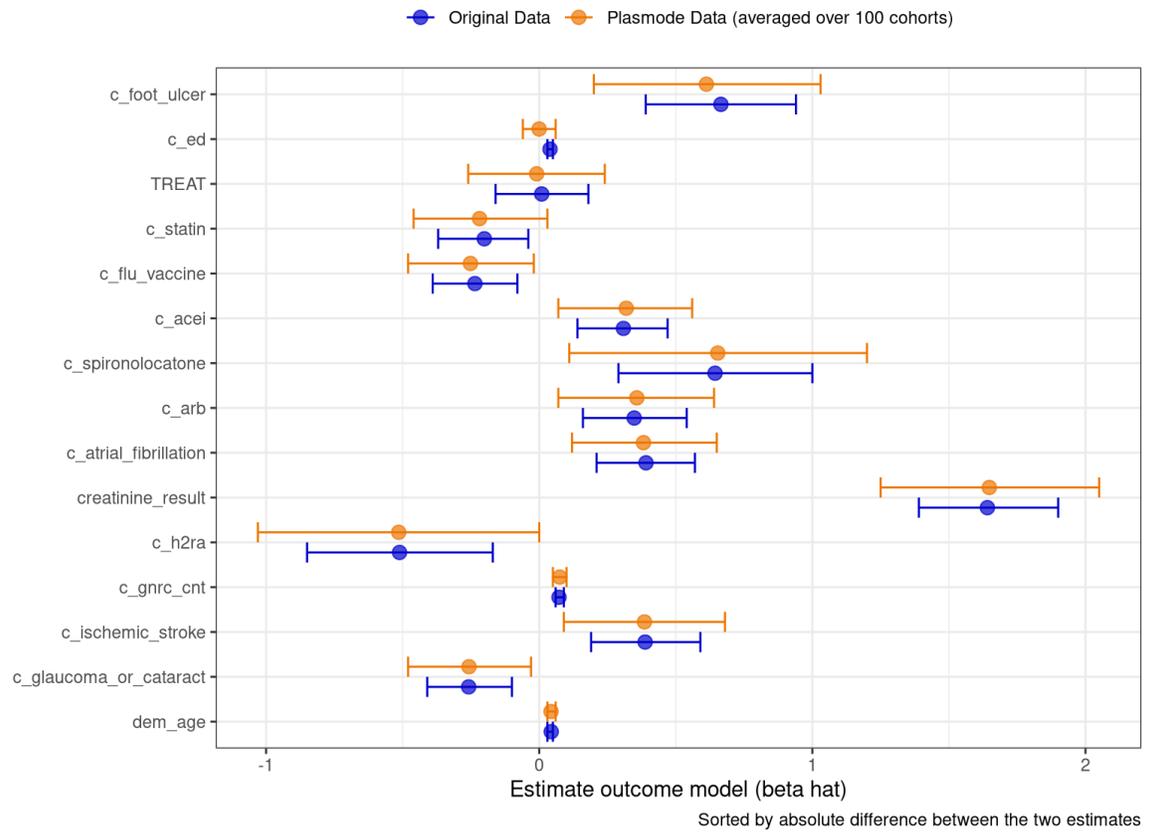

Sorted by absolute difference between the two estimates

Supplementary Figure 3: Overview on # of covariates by considered data dimensions for data generation and the high-dimensional multiple imputation algorithm.

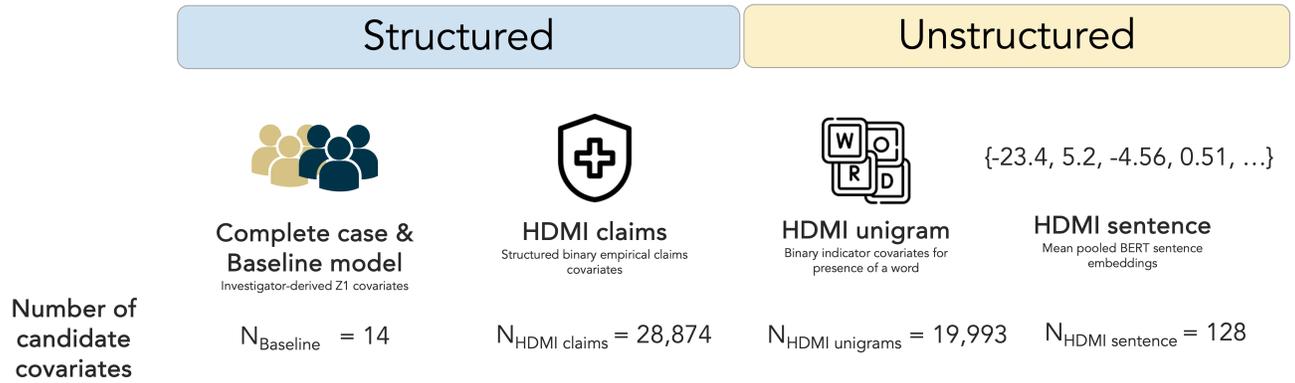

Supplementary Figure 4: Observed crude cumulative incidence rate for acute kidney injury in the eligible complete cohort that was used as a basis to create plasmode cohorts.

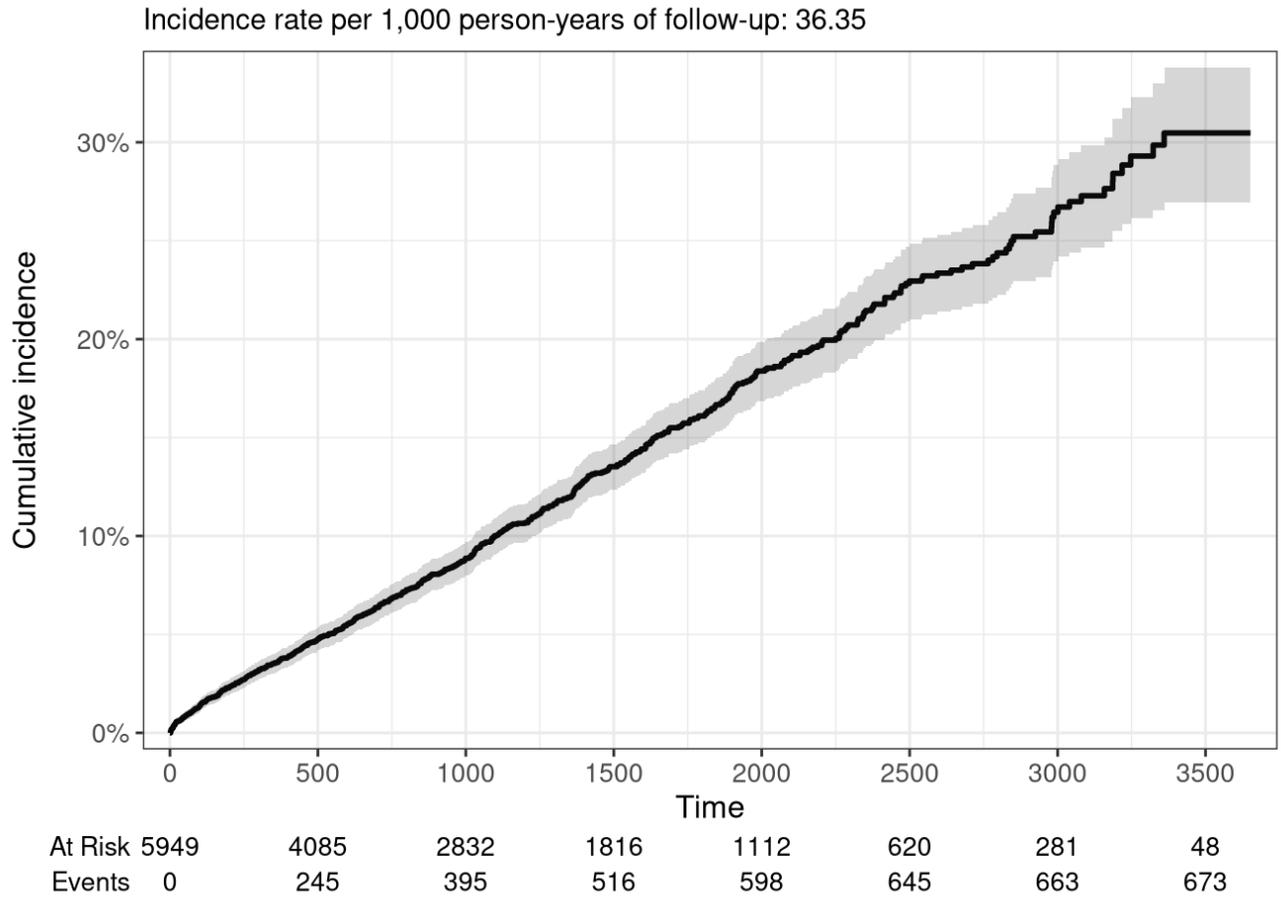

Supplementary Figure 5: Distribution of the mean proportion of encounters captured in EHR data (continuity score, x-axis) among all patients in the eligible complete cohort.

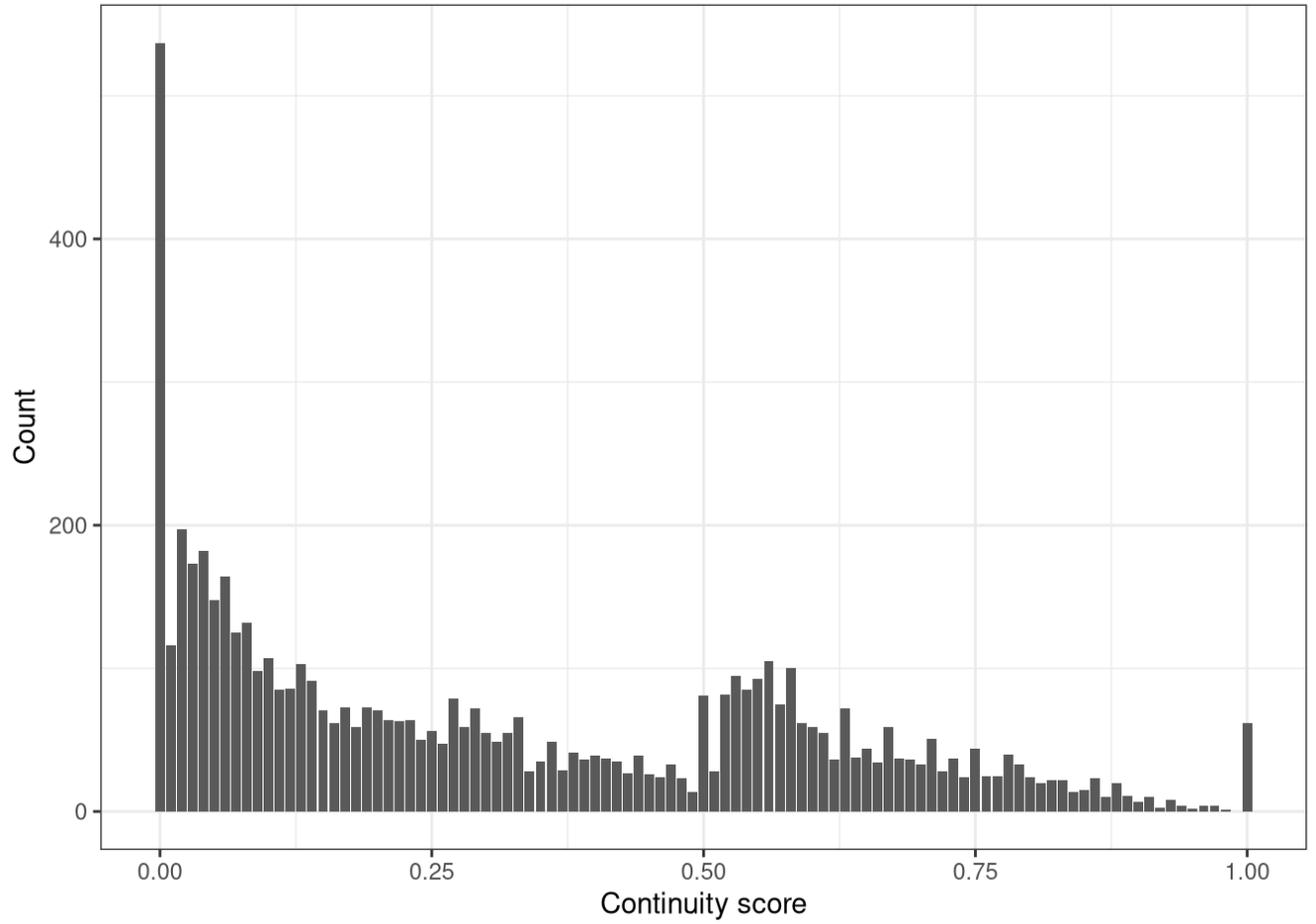